\pdfoutput=1

\documentclass[fleqn,usenatbib]{mnras}

\usepackage{newtxtext,newtxmath}

\usepackage[T1]{fontenc}

\DeclareRobustCommand{\VAN}[3]{#2}
\let\VANthebibliography\thebibliography
\def\thebibliography{\DeclareRobustCommand{\VAN}[3]{##3}\VANthebibliography}

\usepackage{graphicx}
\usepackage{amsmath}

\usepackage{orcidlink}
\usepackage{multirow}

\title[A New Double-Peaked Quasar]{Discovery of a Quasar with Double-Peaked Broad Balmer Emission Lines}
\author[J.H. Terwel \& P.G. Jonker]{
Jacco H. Terwel\orcidlink{0000-0001-9834-3439},$^{1}$\thanks{E-mail: terwelj@tcd.ie (TCD)}
Peter G. Jonker\orcidlink{0000-0001-5679-0695}$^{2,3}$
\\
$^{1}$School of Physics, Trinity College Dublin, The University of Dublin, College Green, Dublin 2, Dublin, Ireland\\
$^{2}$Department of Astrophysics/IMAPP, Radboud University, P.O. Box 9010, 6500 GL Nijmegen, the Netherlands\\
$^{3}$SRON, Netherlands Institute for Space Research, Niels Bohrweg 4, 2333 CA Leiden, the Netherlands\\
}

\date{Accepted XXX. Received YYY; in original form ZZZ}

\pubyear{2022}

\begin{document}
\label{firstpage}
\pagerange{\pageref{firstpage}--\pageref{lastpage}}
\maketitle

\begin{abstract} 
Most massive galaxies contain a supermassive black hole (SMBH) at their center. When galaxies merge, their SMBHs sink to the center of the new galaxy where they are thought to eventually merge. During this process an SMBH binary is formed. The presence of two sets of broad emission lines in the optical spectrum of an active galactic nucleus (AGN) has been interpreted as evidence for two broad line regions (BLR), one surrounding each SMBH in a binary. We modeled the broad Balmer emission lines in SDSS spectra of 373 extreme variability AGNs using one broad and several narrow Gaussian components. We report on the discovery of SDSS J021647.53$-$011341.5 (hereafter J0216) as a double-peaked broad emission line source. Among the 373 AGNs there were five sources that are known double-peaked emission line sources. Three of these have been reported as candidate SMBH binaries in previous studies. We present all six objects and their double-peaked broad Balmer emission lines, and discuss the implications for a tidal disruption event (TDE) interpretation of the extreme variability assuming the double-peaked sources are SMBH binaries. 
\end{abstract}


\section{Introduction}

The current consensus is that most massive galaxies host a supermassive black hole (SMBH, M$_\text{BH}\gtrsim10^6$M$_\odot$) in their nuclei \citep{SMBH_host_coevolution}. One notable group of galaxies contain an active galactic nucleus (AGN) \citep{SMBH_overview}. AGNs are the result of matter falling in towards the central SMBH. This accretion process can release enough energy for the AGN to outshine the entire host.\par

AGNs are divided into different categories depending on their properties. The brightest ones are called quasars, or quasi-stellar objects (QSOs), the less bright ones are Seyfert galaxies. In a type I AGN both broad and narrow emission lines are detected in a rest-frame wavelength ultra-violet to near-infrared spectrum. For type II AGNs on the other hand only narrow emission lines are detected. One explanation for these observed differences is that different orientations of the AGNs allow certain signatures to be detected \citep{unified_models}. One can also make a distinction between radio-loud and radio-quiet AGNs, which is proposed to be caused by the SMBH spin \citep{unification_2D}, although this is still uncertain.\par

Accretion rate changes will result in changes in the AGN's  optical luminosity. Some AGNs have been observed to undergo extreme variability, changing by as much as a magnitude in brightness within a year. A subclass of these extreme variability AGNs have been observed to transition between type I and II. These are called changing look AGNs (CLAGNs) or changing look quasars (CLQs). Various mechanisms have been suggested to explain these changes, like microlensing \citep{microlensing_explanation}, tidal disruption events (TDEs) \citep{TDE_expkanation}, or state changes caused by different mass accretion rates through the disk \citep{magnetic_explanation, Cannizzaro2020}.\par

After a galaxy merger the SMBHs, if one was present in each galaxy prior to the merger, will sink to the new nucleus through dynamical friction \citep{Dynamical_friction} and form an SMBH binary system. This process is the most efficient in reducing the orbits until the SMBHs form a Keplerian system with a separation $a_\text{binary}$, which can be expressed in terms of the mass ratio of the binary $q=m_2/m_1 \leq 1$, its total mass $m_\text{total}=m_1+m_2$, and the velocity dispersion of the bulge stars $\sigma_\star$ as
\begin{equation}
    a_\text{binary} \sim 0.1 \frac{q}{(1+q)^2} \left(\frac{m_\text{total}}{10^6 \text{M}_\odot}\right) \left(\frac{100 \text{km s}^{-1}}{\sigma_\star}\right)^2 \text{pc},
\end{equation}
\citep{Path_to_Coalescence}.
At this point other processes must allow for the transport of binary orbital angular momentum until gravitational radiation becomes the dominant mechanism to transport angular momentum at $\lesssim10^{-2}$ pc (e.g. \citealt{SMBH_grav_waves}). The mechanism(s) responsible for shrinking the separation across this gap is unknown, and the timescale associated with it is highly uncertain. This is generally referred to as the "final pc problem". Several mechanisms have been suggested, like a continuous supply of stars to scatter \citep{final_parsec_scatter}, interactions with a gas disc \citep{final_parsec_discs}, or interaction with a third SMBH leaving a harder SMBH binary while one of the three SMBHs may be ejected \citep{final_parsec_multiple_SMBH}. In order to test these models, observations of SMBH binaries, ideally in various stages of the hardening process, are required to compare against.\par

In the case that both SMBHs in a binary are actively accreting, the observed properties of the AGNs could reveal the binary nature. Before the binary is formed, when two AGNs are completely detached, there are two broad line regions (BLRs) and two narrow line regions (NLRs). As the NLR is typically at a distance of several hunderd pc up to a kpc from the SMBH, these are the first to merge, surrounding the whole system. This means that the emission lines of the NLR can be used to estimate the host galaxy redshift. The BLR, on the other hand, has a typical distance of a few light days to light weeks from the SMBH, and will still orbit an individual SMBH even when it is in the binary. Initially the center of the broad line in the spectrum will be Doppler shifted periodically on the binary orbit. The magnitude of this shift depends on the SMBH velocity, the orbital phase and its inclination, but velocities of several hundreds km s$^{-1}$ are expected \citep{SMBHB_signal_theory}. If both SMBHs have BLRs, the broad lines will for such Doppler velocities (partially) overlap in the spectrum resulting in a complex line shape, possibly showing two peaks or a broad line whose blue and red side are different. If only one SMBH has a BLR, it will result in an offset in wavelength between the broad and narrow emission lines. See \citet{SMBHB_multi_messenger} for further discussion.\par

We report the discovery of double peaked broad emission lines in SDSS J021647.53$-$011341.5 (hereafter J0216). The source was found in our survey of the emission line properties of extreme variability AGNs. Through our survey we also found three previously reported candidate SMBH binaries back, and two sources for which double-peaked emission lines have been reported before, although they were not reported as candidate SMBH binaries before. In Section~\ref{data} we show our selection method and present the procedure used to fit the emission line profile of the broad H~$\alpha$ and H~$\beta$ lines as well as the various emission lines originating in the NLR. Section~\ref{results} shows the best fitting models, which are discussed in Section~\ref{discussion}.

\section{Data}
\label{data}

In this work, the extreme variability quasars and CLAGN catalogs from \citet{Macleod2016}, \citet{Ruan2016}, \citet{Graham2017}, and \citet{Rumbaugh2018} were combined to form a list of 999 unique objects. Spectra for a large fraction of these objects obtained through the Baryon Oscillation Spectroscopic Survey (BOSS) \citep{BOSS1,BOSS2} and/or the Sloan Digital Sky Survey (SDSS) DR12 \citep{SDSS1,SDSS2, SDSSDR12} are available.\par

The original aim of our study was to estimate the central SMBH mass to determine if the SMBHs were too massive for the extreme AGN variability to (easily) be explained as due to a TDE \citep{thesis_JHT}. This estimate is based on single epoch reverberation mapping using the broad H~$\alpha$ and H~$\beta$ lines \citep{Single_epoch_reverberation_mapping}. For this reason, and given the wavelength coverage of the spectra from BOSS and SDSS, two conditions are placed: 1) a source redshift z $\leq1.2$ and 2) a median signal to noise ratio over the spectrum $\geq4$. Out of the 999 extreme variability AGNs 373 met these conditions.

\subsection{Fitting procedure}
\label{fitting_procedure}
The \textsc{lmfit} \textsc{python} package \citep{lmfit} was used for fitting the SDSS spectra of the 373 sources. For the initial fits the H~$\alpha$ and H~$\beta$ regions (called region 1 and 2 respectively) were fitted separately. In region 1 the fit function consisted of a first order polynomial, a narrow and a broad H~$\alpha$ emission line, and the [N~II]$_{\lambda\lambda6548, 6583}$ narrow emission lines. In region 2 the fit function consisted of a first order polynomial, a narrow and a broad H~$\beta$ emission line, and the [O~III]$_{\lambda\lambda4959, 5007}$ narrow emission lines. Each narrow line was only added if it improved the fit significantly. This was determined using the Akaike information criterion (AIC, \citealt{AIC}). For a model with $k$ free parameters and maximum likelihood $L$ it is defined as $\text{AIC} = 2k - 2\text{ln}(L)$, and aims to select the model that best describes the observations while penalizing models with more free parameters. As such, the model with the lowest AIC score describes the data best. The difference in AIC between two models, $\Delta\text{AIC} = \text{AIC}_\text{1} - \text{AIC}_\text{2}$, can be used to asses if one model is favoured over the other. If $|\Delta\text{AIC}| > 10$, the model with a lower AIC is strongly favoured \citep{DeltaAIC}. All emission lines in a given model were assumed to have a Gaussian shape and to have the same redshift. Additionally, all narrow lines in region 1 as well as [O~III]$_{\lambda\lambda4959, 5007}$ were assumed to have the same Full Width at Half Maximum (FWHM). The wavelength error of SDSS and BOSS are calibrated to be $< 5$ km s$^{-1}$. As the narrow and broad emission lines have a typical FWHM of several hundred km s$^{-1}$ and several thousand km s$^{-1}$ respectively, this error has been neglected.\par

The results of these fits can be put into three broad categories: 1) The best-fitting model (the emission lines associated with a single SMBH, its BLR, and NLR) describes the data well. 2) The best-fitting model to the data provides an unsatisfactory high $\chi^2_\text{red}$ ($\chi^2_\text{red}\geq30$). Inspection of the fit and data shows that this is caused by the absence of a detectable broad line in the spectrum. 3) The broad emission line cannot be fit well by a single Gaussian. For instance, the broad emission line appears to have multiple peaks, or the blue and red slope of the broad component are different.\par

After the objects whose broad features could not be reproduced with a single broad line were selected, they were fitted again using an updated procedure. First, the shape of the [O~III]$_{\lambda5007}$ emission line is fitted to serve as a model for all narrow lines. Gaussian, Lorentzian, and Voigt profiles were considered. This sets the shape, redshift, and FWHM for all narrow lines in both regions and fit functions. In region 1 the [S~II]$_{\lambda\lambda6716, 6730}$ doublet is included in the description of the narrow emission lines. Next, both regions are fitted with a first order polynomial continuum, narrow lines with only their height as a free parameter, and one or two broad Gaussian lines for which the six parameters are allowed to float freely.

\section{Results}
\label{results}

\begin{table*}
	\centering
	\caption{The objects with double-peaked broad Balmer emission lines. The values for $z$ are found by fitting the narrow [O~III]$_{\lambda5007}$ emission line and may differ slightly from the values quoted by SDSS. Columns 3 and 4 show the observation date of the used spectrum and the instrument used. Columns 5 and 6 show the $\Delta\text{AIC} = \text{AIC}_\text{1 broad} - \text{AIC}_\text{2 broad}$ scores. Columns 7 to 10 show the wavelength difference between the fitted broad lines and the emission line wavelength at rest in the source rest frame as $\Delta\lambda = \lambda_\text{broad} - \lambda_\text{line}$. The best-fit narrow line profile is given in column 11. J0216 is on the top row, while the other sources are listed in order of increasing right ascension.}
	\label{obj_specs}
	\begin{tabular}{lcccccccccr}
		\hline
		Name &  & Observed & Instrument & \multicolumn{2}{c}{$\Delta\text{AIC}$} & \multicolumn{2}{c}{$\Delta\lambda$ region 1 (\AA)} & \multicolumn{2}{c}{$\Delta\lambda$ region 2 (\AA)} & narrow line\\
		SDSS J & $z$ & (UT) & & region 1 & region 2 & blue & red & blue & red & profile\\
		\hline
		021647.53$-$011341.5 & $0.41993(2)$ & 10-10-2010 & BOSS & 377 & 225 & -13(1) & 20.9(8) & -13(1) & 21(1) & Lorentzian\\ 
		000710.01+005329.0 & $0.31611(3)$ & 02-09-2002 & SDSS & 354 & 229 & -59(1) & 25(2) & -45(1) & 41(5) & Voigt\\ 
		004319.74+005115.4 & $0.30823(3)$ & 07-09-2000 & SDSS & 338 & 170 & -83(2) & 58(4) & -58(2) & 51(3) & Voigt\\ 
		021259.59$-$003029.4 & $0.39448(2)$ & 26-09-2000 & SDSS & - & 125 & - & - & -39(2) & 34(4) & Voigt\\ 
		022014.58$-$072859.2 & $0.21369(2)$ & 10-09-2001 & SDSS & 387 & 67 & -18(1) & 93(3) & -39(2) & 30(7) & Lorentzian\\ 
		222024.58+010931.3 & $0.21227(8)$ & 19-08-2001 & SDSS & 112 & 69 & 4(1) & 20(3) & -18(3) & 30(10) & Gaussian\\ 
		\hline
	\end{tabular}
\end{table*}

Table \ref{obj_specs} shows the SDSS name of each object with double-peaked broad Balmer emission lines, its redshift as found by fitting the narrow [O~III]$_{\lambda5007}$ emission line, and the date at which the spectrum was observed. For the rest of the paper, each object will be referred to by the first four digits of its right ascension, e.g., J2220. The AIC of the fit with one broad Gaussian is compared directly to the AIC of the fit with two broad Gaussians to assess the impact of adding an additional broad emission line. We define $\Delta\text{AIC}$ in columns 5 and 6 of Table \ref{obj_specs} as the AIC of the fit with one broad Gaussian minus the AIC of the fit with two broad Gaussians. Columns 7 to 10 show the wavelength difference between the fitted broad lines and the emission line wavelength at rest in the source rest frame. Lastly, column 11 shows the best-fitting function used to describe the narrow lines.\par

\begin{figure*}
	\includegraphics[width=0.9\textwidth]{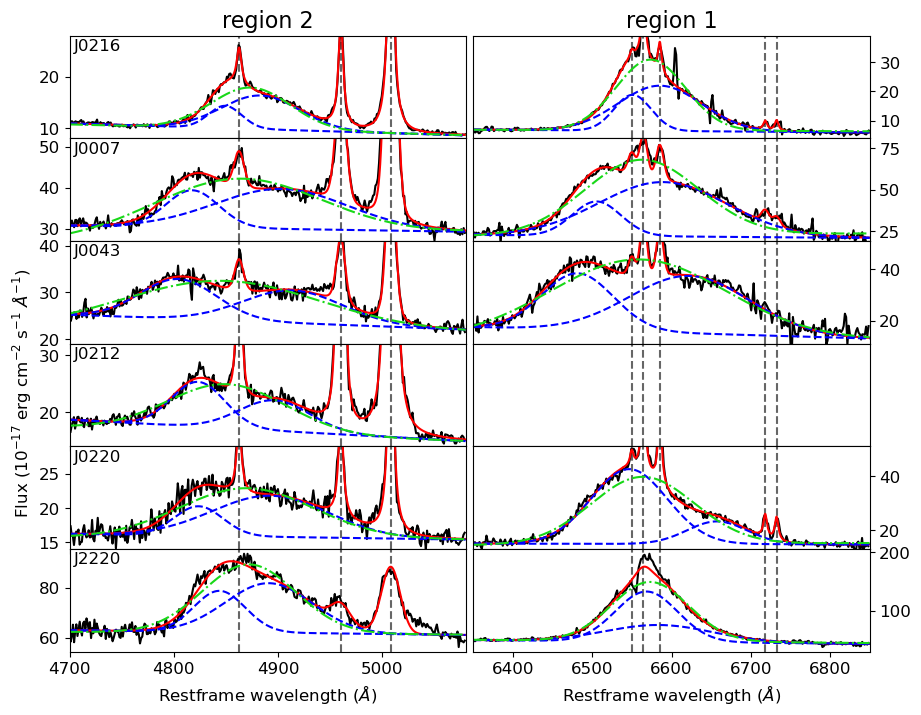}
    \caption{The spectra for the six objects where the observed broad Balmer emission line profiles cannot be described well with a single Gaussian. The observed spectrum is shown in black, with the best-fitting double broad line model overlaid in red. The individual broad lines are shown by the dashed blue lines. For comparison the best fitting single broad line model is shown using the dot-dashed green line. The vertical dashed grey lines give the locations of narrow emission lines. From left to right the fitted emission lines are: H~$\beta$, the [O~III]$_{\lambda\lambda4959, 5007}$ doublet, the [N~II]$_{\lambda\lambda6548, 6583}$ doublet with H~$\alpha$ in the middle, and the [S~II]$_{\lambda\lambda6716, 6730}$ doublet. Region 1 of J0212 is not observed as it is shifted outside the wavelength range accessible by SDSS due to the high $z$ of the source.}
    \label{fitted_models}
\end{figure*}

Figure \ref{fitted_models} shows regions 1 \& 2 for the six sources with evidence for the presence of double-peaked broad Balmer lines fitted by the double broad line model (red line). The two broad lines are shown separately (blue dashed line), as well as the best fit for the single broad line model for comparison (green dot-dashed line). The positions of the narrow emission lines are shown with vertical dashed lines. Next, we will discuss the spectra of the 6 double-peaked Balmer line sources.

\subsection{Newly discovered candidate SMBH binary: J0216}
As can be seen in Figure~\ref{fitted_models}, the broad emission line complex has a steeper blue than red wing. The attempt to fit the line complex with a fit-function containing only one broad emission line results in a poor fit, with a $\chi^2_\text{red}$ of 17.7 and 14.0 for region 1 and 2, respectively. The model has 10 degrees of freedom (DOF) in region 1, and 8 DOF in region 2. A model with two broad Gaussian components reduces the $\chi^2_\text{red}$ to 4.5 and 7.2 for region 1 and 2, respectively ($\Delta\text{AIC} =$ 377 and 225). These models have 3 more DOF compared to the single broad Gaussian models.\par

There seem to be two narrow lines that are not modelled in region 1, the first is around 6600 \AA~and the second around 6650\AA. These are the residuals of two strong sky lines. The data points associated with these sky lines have large error bars and do not significantly influence the best-fitting parameters of the fit-function.

\subsection{Known double-peaked emission line sources}
For the other five sources the fit-function containing two broad lines have a significantly lower AIC score compared to the fit-function composed of a single broad line (see Table~\ref{obj_specs}). This means that the fit-function with two broad lines is strongly favoured. As can be seen in Figure~\ref{fitted_models}, the best fits using a single broad Gaussian to fit the Balmer emission line (green dashed-dotted line) leave large parts of the emission complex unfitted. Furthermore, the FWHM of such a simple broad line fit would imply an extreme BH mass assuming that the emission originates from the BLR. These problems are solved by using two broad Gaussian emission lines.

\section{Discussion}
\label{discussion}

In a project designed to measure the SMBH mass in CLAGNs through fitting the broad Balmer emission line profiles we discovered double-peaked broad Balmer lines in 1 extreme variability quasar (J0216) and rediscovered 5 previously known sources with complex broad Balmer emission lines \citep{low-ion_double-peaks, double-peaked_low_ion_4_obj, Double-peaks_Mbh}. For these six objects, a fit-function that includes a second broad emission line to describe the broad Balmer emission complexes have significantly lower AIC scores when compared with a fit-function composed of a single broad line to describe either the broad H~$\alpha$ or H~$\beta$ line. Three of the known sources (J0043, J0212, and J0220) have been reported as candidate SMBH binaries before \citep{SMBHB_orbit_timing, SMBHB_Doppler_boost, sys_search_SMBHB, sys_search_SMBHB_2, sys_search_SMBHB_3}. J0007 is included in the search for AGN pairs in \cite{AGN_pairs}, but is not reported to be one. An AGN pair can be observed before or during a galaxy merger, while the SMBH binary is created after the merger when the nuclei have sunk to the center of the merged galaxy.

Several alternative models have been suggested to explain the origin of double-peaked emission-line profiles. \cite{J2220_absorption} argue that part of the light coming from the BLR in J2220 is absorbed, creating non-Gaussian broad line profiles. \cite{alt_outflow} suggested that a bipolar outflow coming from jets interacting with the gas immediately around the AGN could create similar line profiles. \cite{alt_anisotropic1} and \cite{alt_anisotropic2} suggested a spherically symmetric BLR being illuminated anisotropically by the accretion disk could cause the broad Balmer emission lines to appear double peaked. \cite{alt_ion_torus} suggested that the double peaks come from the outer regions of the accretion disk, which is illuminated by a thick ion torus closer to the central SMBH. This is shown to fit the data of nearby double-peaked type I AGN in \cite{Double_peaks_from_disk}. All of these models are discussed in \cite{alt_discussion}, and they favour the accretion disk illuminated by a thick ion torus scenario. They provide several reasons against the other models: e.g., the line emitting gas is not in the same region as the bipolar outflow, and the short lifetime of the configuration required in the anisotropically illuminated BLR scenario.\par

One of the main criticisms against the SMBH binary model is that it is assumed that both SMBHs orbit the center of mass, and for each SMBH there is a component of this movement in our line-of-sight causing the central wavelengths of the broad lines to be offset from their narrow line equivalent. The magnitude of the offsets depends on the phase and inclination of the system, and the broad lines should move from the blue side to the red side of the narrow line and back in a sinusoidal pattern (see figure 4 of \citealt{SMBHB_signal_theory}). Even if the period of the orbit has a timescale of centuries, one would expect to see changes over time in the position of the broad lines over the span of one or several decades \citep{alt_discussion}. \cite{Rad_vel_test_SMBHB_scenario} tested this hypothesis for 13 low-$z$ double-peaked AGN with time series spanning up to nearly 40 years, but found the SMBH masses required to be several orders of magnitude larger than inferred from e.g. reverberation mapping. \cite{SMBHB_kinematics_observability} also showed that to detect an SMBH binary through kinematic signatures in their broad lines an extreme mass ratio and specific separation is needed for the SMBH velocities to be observable while the BLRs are separated.
Related to this is the need for the center of mass of the SMBH binary to be at rest with respect to the redshift of the host galaxy. For Mrk 668, one of the first proposed candidate SMBH binaries \citep{1st_candidates}, the parameters of the broad Balmer emission lines were monitored for more than a decade. It was found that the radial velocity is not sinusoidal \citep{mrk668_non_sinus}, and the center of mass is at a different redshift from the host galaxy \citep{mrk668_wrong_barycenter}.\par

Except for region 1 of J2220 all modelled regions have one broad line clearly on each side of their narrow line counterparts, as expected in the SMBH binary scenario. \cite{SMBHB_orbit_timing} found a periodic signal in the photometry of J0043 in data from the Palomar Transient Factory (PTF, \citealt{PTF}) and Catalina Real-Time Transient Survey (CTRS, \citealt{CRTS}). \cite{SMBHB_Doppler_boost} showed that the aperiodic parts of the variability of this object are consistent with relativisic Doppler boosting in an SMBH binary. \cite{sys_search_SMBHB} attempted to measure shifts in the broad H~$\beta$ profile of their targets using additional spectra taken in 2009. They found a lower and upper limit on the shift in J0212, but were unable to quantify the shift in J0220, as the H~$\beta$ profile varied substantially between the observations. \cite{sys_search_SMBHB_3} tried to improve on their radial velocity measurements, but were unable to in the case of these two sources again due to the variability of the H~$\beta$ profile.\par

When comparing the separation between the two broad lines in the best-fitting models it can be seen that the separation in J0216 is among the lowest, with only the broad lines in region 1 of J2220 being lower (see Table~\ref{obj_specs}). Figure~\ref{fitted_models} also shows that for all other sources there is at least one broad line that is significantly broader than the broad lines in J0216. In the SMBH binary interpretation this can be explained as a result of the mass of the SMBHs. A broader broad line suggest a faster moving BLR around a more massive SMBH. This is used in e.g. single epoch reverberation mapping SMBH mass estimations \citep{Single_epoch_reverberation_mapping}. A similar reasoning can be used for the separation between the broad lines. The velocity of the individual SMBHs is expected to be higher if the components are more massive. For a similar inclination and phase, broader broad lines that are more separated are expected to be seen in more massive SMBH binaries. These separations are also expected to be consistent between region 1 and 2. This is only observed in J0216 and J0007.\par

In the case where the extreme variability of these 6 objects is caused by a TDE, there is an upper limit on the SMBH mass. This is because the event horizon of an SMBH inceases linearly with mass, while the tidal radius  depends on the cube root of the SMBH mass. The maximum mass for an SMBH TDE, called the Hills mass $M_\text{H}$ \citep{Hills}, is $M_\text{H} = 6.2 \times 10^8 r_\text{IBCO}^{-3/2}(a) M_\star^{-1/2} R_\star^{3/2} M_\odot$ with $r_\text{IBCO}(a)$ the innermost bound circular orbit depending on SMBH spin $a$, and $M_\star$ and $R_\star$ the stellar mass and radius measured in solar mass and radius respectively \citep{Hills_gen_rel}. \cite{Mvir_estimates} estimated the virial SMBH mass of 4 of these 6 objects using the H~$\beta$ and Mg~II lines, assuming a single Gaussian broad component for each line. In those 4 cases the estimated mass was close to or above the Hills mass. A SMBH binary can explain the features of the broad lines while keeping at least one SMBH below the Hills mass, allowing for a TDE interpretation of the observed extreme variability.\par

\section{Conclusion}
We have found an extreme variability quasar with double-peaked broad Balmer emission lines: J0216. We showed that the broad Balmer emission observed in the spectrum of this extreme variability quasar cannot be described well using a single Gaussian. Including a second broad Gaussian significantly reduced the $\chi^2_\text{red}$ value of the best fit profiles. We also recovered the already known double peaked broad Balmer emission line profiles for five extreme variability quasars, showing the reliability of our method. Three of these have been reported as SMBH binary candidates before. A SMBH binary opens the door to a TDE interpretation of the observed extreme variability in these objects by allowing at least one of the SMBHs to be below the Hills mass. However, the SMBH binary model faces several challenges, and is disfavoured compared to, e.g., an outer disk origin of the double-peaked emission.

\section*{Acknowledgements}
Funding for SDSS-III has been provided by the Alfred P. Sloan Foundation, the Participating Institutions, the National Science Foundation, and the U.S. Department of Energy Office of Science. The SDSS-III web site is http://www.sdss3.org/.
SDSS-III is managed by the Astrophysical Research Consortium for the Participating Institutions of the SDSS-III Collaboration including the University of Arizona, the Brazilian Participation Group, Brookhaven National Laboratory, Carnegie Mellon University, University of Florida, the French Participation Group, the German Participation Group, Harvard University, the Instituto de Astrofisica de Canarias, the Michigan State/Notre Dame/JINA Participation Group, Johns Hopkins University, Lawrence Berkeley National Laboratory, Max Planck Institute for Astrophysics, Max Planck Institute for Extraterrestrial Physics, New Mexico State University, New York University, Ohio State University, Pennsylvania State University, University of Portsmouth, Princeton University, the Spanish Participation Group, University of Tokyo, University of Utah, Vanderbilt University, University of Virginia, University of Washington, and Yale University.

This is a pre-copyedited, author-produced PDF of an article accepted for publication in MNRAS following peer review. The version of record is available at: \url{https://academic.oup.com/mnrasl/advance-article-abstract/doi/10.1093/mnrasl/slac026/6551317}

\section*{Data Availability}

The data used in this letter is available through the SDSS DR12 Database \citep{SDSSDR12}. The scripts used to fit the data and the final fit function parameters are available on Zenodo after paper acceptance \citep{scripts+data}.

\bibliographystyle{mnras}
\bibliography{main}

\bsp
\label{lastpage}
\end{document}